# Feasible Nanometric Magnetoresistance Devices

Oded Hod[1], Roi Baer[2,♦], and Eran Rabani[1,♦]

[1]School of Chemistry, The Sackler Faculty of Exact Science, Tel Aviv University, Tel Aviv 69978, Israel

[2]Department of Physical Chemistry and the Lise Meitner Center for Quantum Chemistry, The Hebrew University of Jerusalem, Jerusalem 91904, Israel.

Abstract

The electrical conductance through a ring is sensitive to the threading magnetic flux. It contains a component that is periodic with an Aharonov-Bohm (AB) period equal to the quantum flux. In molecular/atomic loops on the nanometer scale, encircling very small areas, the AB period involves unrealistically huge magnetic fields. We show that despite this, moderate magnetic fields can have a strong impact on the conductance. By controlling the lifetime of the conduction electron through a pre-selected *single state* that is well separated from other states due to the quantum confinement effect, we demonstrate that magnetic fields comparable to one Tesla can be used to switch a nanometric AB device. Using atomistic electronic structure calculations, we show that such effects can be expected for loops composed of monovalent metal atoms (quantum corals). Our findings suggest that future fabrication of nanometric magnetoresistance devices is feasible.

Understanding nanoscale electronic devices is intertwined with the ability to control their properties[1-10]. In mesoscopic systems an effective control of conductance in loops is obtained by the "turn of the electromagnet knob"[11-16] exploiting the Aharonov-Bohm (AB) effect[17]. This has lead to development of micronic AB interferometers[18-22]. While large magnetoresistance has been demonstrated for molecular systems based on Zeeman[23] splitting or the Kondo effect[24], in the nanoscale however, it is widely accepted that AB magnetoresistance devices do not exist. This is because unrealistic huge magnetic fields are required to affect conductance in loops encircling very small areas.

In this letter we present the necessary physical principles for constructing a *feasible* nanoscale magnetoresistance device based on the AB interferometer. We demonstrate that by controlling the lifetime of the conduction electron through a pre-selected single state, moderate magnetic

♦ To whom correspondence should be sent: R.B. (roi.baer@huji.ac.il) or E.R. (rabani@tau.ac.il).

fields can switch the conductance of a nanometer atomic corral. Our findings suggest for the first time that future fabrication of nanometric devices with positive magnetoresistance is feasible.

To present the basic principles for our nanometric device, let us first consider a well-known model of the AB interferometer[25,26], consisting of a 1D continuum loop connected to two wires as described in Figure 1. The loop of area $A$ is placed in a perpendicular magnetic field $B$, and the threading magnetic flux is therefore $\Phi = AB$. Since we are interested in nanometric size devices, the model does not address the effects of disorder, which in micrometric devices plays an important role[27].

The transmission probability $T(\Phi)$ that a conduction electron of energy $E_k$ originating in the left wire passes through the loop emerging through the right wire can be calculated exactly[28]. The result for the transmission $T(\Phi)$ can be cast in terms of two independent parameters[29]: the junction scattering amplitude $\varepsilon$ defined in Figure 1, and the spatial phase angle $\theta_k = kL/2$. Here, $k = \sqrt{2\mu_e \hbar^{-2} E_k}$ is the wave vector, $L$ the circumference of the loop, and $\mu_e$ is the electron mass.

The correct combination of $\varepsilon$ and $\theta_k$ is crucial for the mechanism discussed now. In Figure 1, we analyze the effects of these parameters on the transmission probability $T(\Phi)$. The principal effect of changing the incoming electron energy (determined by $\theta_k$) is shown in the middle panel of Figure 1 for a given value of $\varepsilon$. As expected, $T(\Phi)$ is periodic with period equal to $\Phi_0 = h/e$. We find that within each period, $T(\Phi)$ has a symmetric structure around $\Phi/\Phi_0 = \frac{1}{2}$, characterized by a double peak. This structure is caused by resonance transmission through the energy levels of the loop. The position of the transmission peaks can be shifted to low magnetic fields by adjusting the kinetic energy ($\theta_k$) of the conduction electron.

In order to achieve a switching capability at feasible magnetic fields, namely at low values of $\Phi$, it is essential to *reduce the width* of the transmission peaks. This can be done by increasing the lifetime of the conduction electron on the ring. In Figure 1 (right panel), we show the dependence of the width of the peaks on the amplitude $\varepsilon$ for entering (leaving) the ring at the left (right) junction. As $\varepsilon$ is decreased from its maximal value of $1/\sqrt{2}$, the electron lifetime on the ring increases, and the transmission peaks become narrow.

In summary, we find that the value of $\theta_k$ controls the *location* of the $T(\Phi)$ maxima, while the value of $\varepsilon$ controls the *width* of its peaks. By carefully selecting the value of these parameters, it is possible to shift the maxima of $T(\Phi)$ to very low magnetic fields, while at the same time dramatically increasing the sensitivity to the magnetic field. Thus, despite the fact that the AB period involves huge magnetic fields (hundreds of Tesla), the mechanism we suggest here can be used to tune the system to respond to magnetic fields of the order of 1 Tesla.

Now the question shifts to the plausibility of assembling such a molecular device. And if so, what are the realistic properties that can be expected. In a realistic system, complications may arise from atomistic disorder and inhomogeneous broadening resulting from the existence of high density of conduction states[30].

In order to check the proposed ideas on a realistic system, we have designed a model of a nanometric AB interferometer composed of Cu atoms arranged in a corral on a metal oxide surface. An illustration of the system is shown in Figure 2, where a ring of 40 Cu atoms is connected to atomic-Cu wires. All atoms are separated by a distance of $R_B = 2.35 \text{ Å}$. An experimental realization of this setup can be achieved using scanning tunneling microscopy (STM) techniques[31-33].

To calculate the conductance we developed a Magnetic Extended Hückel Theory. A similar approach has been used extensively to study conductance in molecular systems[34]. Within this approach, each Cu atom donates ten *d*-electrons and one *s*-electron, and its valence *s*, *p* and *d* orbitals are explicitly considered in the Hamiltonian. The magnetic field **B** is assumed homogeneous in the $z$-direction. We use a gauge-invariant atomic orbital basis[35,36] and calculate the Hamiltonian matrix within the Pople approximation[37]. A gate voltage effect is simulated by adding $eV_g$ to the energies of the atomic orbitals on the ring atoms only.

Conductance is computed using the Landauer formalism[38], which relates the conductance to the transmittance through the molecular system:

$$g = g_0 \frac{\partial}{\partial V} \int [f_L - f_R] T(E) dE, \qquad (1)$$

where $g_0 = 2e^2/h$ is the quantum conductance, $f_{L/R}(E) = \left[1 + e^{\beta(E-\mu_{L/R})}\right]^{-1}$ is the Fermi-Dirac distribution in the left/right lead, $\beta = 1/k_BT$ is the inverse temperature, and $\mu_{L/R}$ is the chemical potential of the left/right lead. $T(E)$ is the transmittance given by[39] $T = 4tr\left\{G^\dagger \Gamma_L G \Gamma_R\right\}$, where $\Gamma_L$ ($\Gamma_R$) are imaginary absorbing potentials inside the left (right) lead representing the imaginary part of the self-energy $\Sigma$, and $G(E) = \left[E - H + i(\Gamma_L + \Gamma_R)\right]^{-1}$ is the appropriate Green's function (we assume the real part of $\Sigma$ is $0$). For additional details, see ref. [29].

In Figure 3 we show the conductance for the Cu corral as a function of the magnitude of the magnetic field $B$ and the gate voltage $V_g$. Two systems are considered containing $4N$ and $4N+2$ (with $N=10$) ring atom, respectively. The common features observed for both systems are: (a) a large magnetic field ($\sim 500-600$ Tesla) is required to complete an AB period. (b) The conductance peaks (red spots) shift with $V_g$. The latter effect is analogous to the shift of peaks seen in Figure 1 (middle panel) where the kinetic energy of the conductance electron is varied (via $\theta_k$). Here, we control the kinetic energy of the conductance electron by determining the molecular orbitals through which conductance takes place. To a good approximation the molecular orbital energy (at $B=0$) is given by an effective mass model $E_m \approx h^2 m^2/2\mu^* L^2$, $m = 0, \pm 1, \ldots$ where $\mu^*$ is the effective mass. Thus, by changing the gate potential $V_g$ we select a molecular orbital with an approximate momentum $hm/L$ through which conductance takes place.

In a half-filled conduction band (such as the Cu s-band considered here), the Fermi wavelength equals to four bond lengths. Thus symmetric loops can be classified into two groups, those containing $4N$ and $4N+2$ atoms. An approximate condition for maximal conductance is given by the following relation between electron wavelength $\lambda$ and loop circumference $L$:

$$L = \lambda\left(m + \Phi/\Phi_0\right). \tag{2}$$

In loops containing $4N$ atoms $\lambda m = L$ and the conductance peak is obtained at zero magnetic fields for $V_g = 0$, as can be seen in Figure 3 (left panel). On the other hand, in $4N+2$ loops, a field corresponding to $\Phi \approx \Phi_0/2$ is needed to satisfy the condition of Eq. (2) at $V_g = 0$, war-ranting a large magnetic field, as seen in Figure 3 (right panel). The condition of Eq. (2) is not exact in this system because of the existence of other energy levels and the broadening due to temperature.

As mentioned, the gate voltage allows control of the location of the maximal conductance. In particular, it can be used to shift the maximal conductance to zero magnetic fields similarly to the control achieved by varying $\theta_k$ in the analytical model. The value of $V_g$ at which this is achieved is different for the two prototypical system sizes considered, and depends on the Fermi wavelength and on the circumference of the loop.

The next step is to control the width of the conductance resonance as a function of the magnetic field. In the analytical model, this was done by reducing the transmission amplitude $\varepsilon$. In the molecular system, this can be achieved by increasing the distance $R_c$ between the edge lead atom closest to the ring and the ring itself (see Figure 2 for an illustration). Alternatively, one can also introduce an impurity atom at the junctions between the leads and the ring. However, for quantum corrals, the former approach seems more realistic.

In Figure 4, the conductance as a function of magnetic field is depicted for several values of $R_c$ for the two generic system sizes. For each system, a proper gate potential is chosen to ensure maximal conductance at $B=0$. As $R_c$ is increased the switching response to the magnetic field is sharpened. At the highest $R_c$ studied, we achieve a switching capability on the order of a *single Tesla*, despite the fact that the AB period is comparable to $500-600$ Tesla.

The above calculations assume a low temperature of 1K; however, the effects hold even at higher temperatures. The temperature $T$ must be low enough to resolve the magnetic field splitting of energy levels and must satisfy $k_B T < \left[4\pi\hbar^2/\mu R_B D\right]\left(\Phi/\Phi_0\right)$, where $D$ is the diameter of the corral, $\mu$ is the effective mass of the electron (in the Cu-Cu corral, $\mu \approx \mu_e$) and as before $R_B$ is the inter-atomic distance. For the studied Cu corral, switching at 1 Tesla, this leads to $T < 30K$.

The realization of a nanometric AB interferometer described above is not limited to the case of Cu atom corrals. In fact, calculations on other monovalent atom systems yield similar quantitative results, and we expect that the approach be valid for heavy metal atoms such as gold. We note that the manipulation of gold atoms on a metal oxide surface has recently been demonstrated with sub-nanometer scale control over the resulting structure[32]. We have also carried out calculations on a more complex system involving two conduction channels, such as a ring composed of carbon atoms (poly-acetylene). The control of the conductance for this system is somewhat more involved; however, a similar qualitative picture emerges.

The physics of the nanometric magnetoresistant device is different from its mesoscopic counterpart. In micrometric interferometers the magnetic field typically increases the conductance due to weak localization, resulting in negative magnetoresistance[15,27]. In contrast, at the nanoscale, disorder can be easily suppressed and positive magnetoresistant behavior emerges. Furthermore, it seems highly unlikely that the effects discussed here can be observed at the micrometric scale because the low magnetic flux implies unrealistically small magnetic fields and the small level spacing in the micrometer interferometer dictates a very low temperature (0.01K) and at such temperatures strong localization may prohibit conductance all together.

Summarizing, we have shown that in spite of its small size, magnetic switching can be achieved in nanometric devices. The essential procedure is to weakly couple the interferometer to the leads, creating a tunneling resonance junction. Thus, conductance is possible only in a very narrow energy window. The resonant state, is tuned by the gate potential, such that at B=0 transmission is maximal. The application of a relatively small magnetic field shifts the interferometer level out of resonance and conductance is strongly reduced.

**Acknowledgements** This research was supported by the Israel Science Foundation and by the US-Israel Binational Science Foundation.

Figures Captions:

Figure 1: (Left panel) Schematics of a circular AB ring-shaped system. The magnetic flux is $\Phi$, measured in units of the quantum flux $\Phi_0 = h/e$. The ring parameters are chosen such that an incoming wave from the left is reflected with amplitude c and enters the top (bottom) branch with amplitude $\varepsilon$ ($2\varepsilon^2 + c^2 = 1$). Middle panel: The transmission probability $T(\Phi)$ for various spatial phases $\theta_k$ at $\varepsilon = 0.3$. Right panel: $T(\Phi)$ for various values of the probability $\varepsilon$ at $\theta_k = 0.5$.

Figure 2: The Cu atom corral. The ring diameter is $\sim 3$ nm. Also shown is the distance $R_c$ at the contact.

Figure 3: Conductance as a function of magnetic field and gate voltage at $T = 1K$ for a ring of 40 (left) and 42 (right) Cu atoms (~3 nm diameter). Rainbow color code: Red corresponds to $g = g_0$ and purple to $g = 0$.

Figure 4: The conductance of $40$ (left) and $42$ (right) Cu atom-corrals at $T = 1K$ as a function of the magnetic field and the contact bond length $R_c$. The AB period is $600$ (left) and $540$ (right) Tesla. The gate potential is $0$ (left) and $-0.132 V$ (right).

Fig 1   Oded Hod, Roi Baer, and Eran Rabani

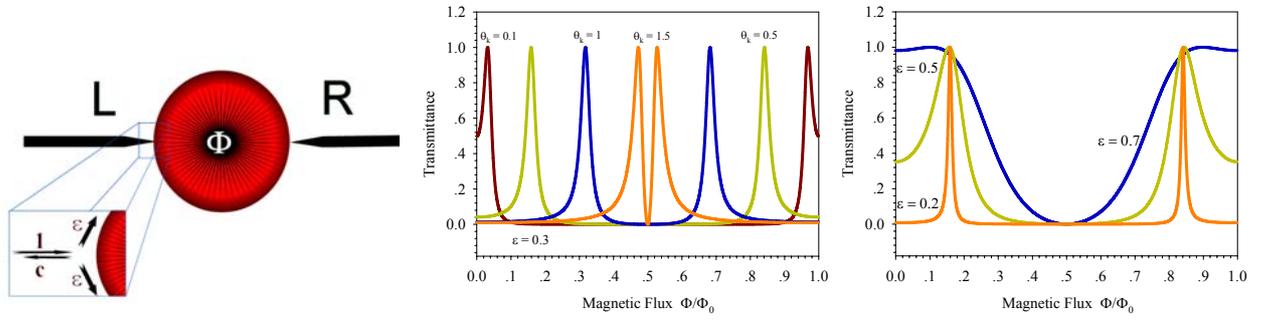

Fig. 2   Oded Hod, Roi Baer, and Eran Rabani

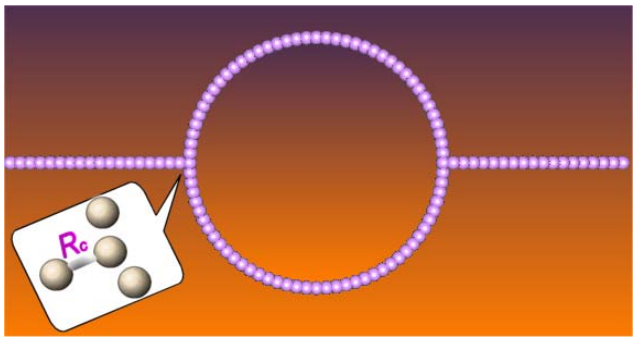

Fig. 3  Oded Hod, Roi Baer, and Eran Rabani

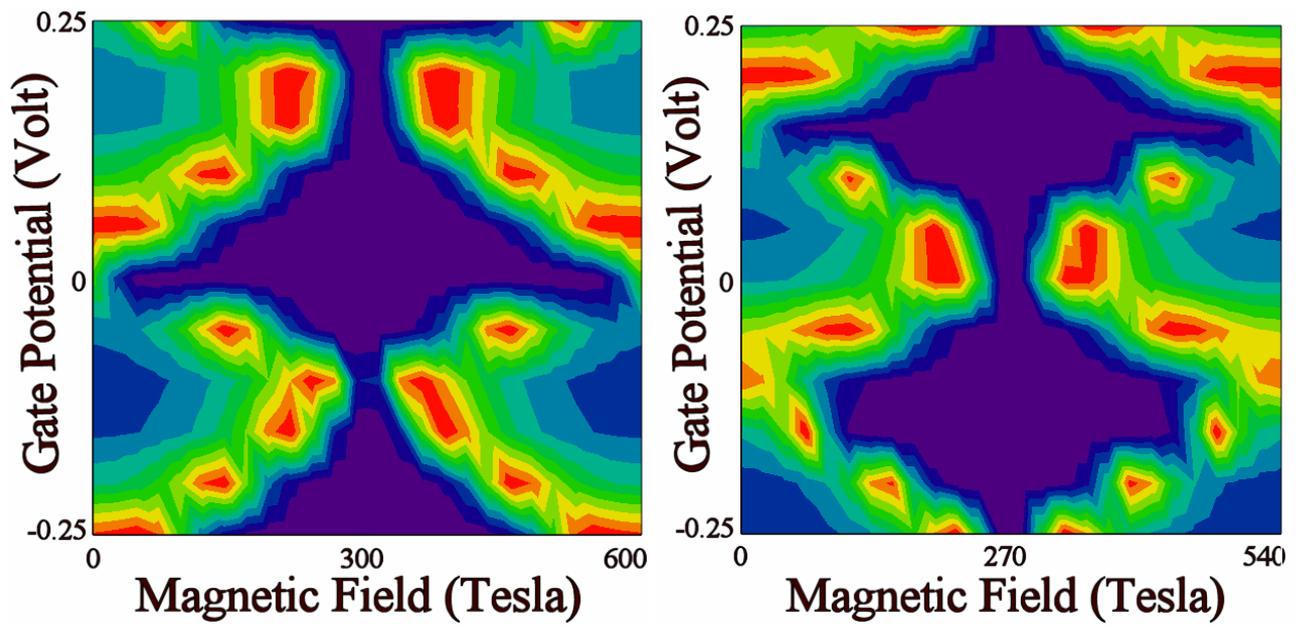

Fig 4  Oded Hod, Roi Baer, and Eran Rabani

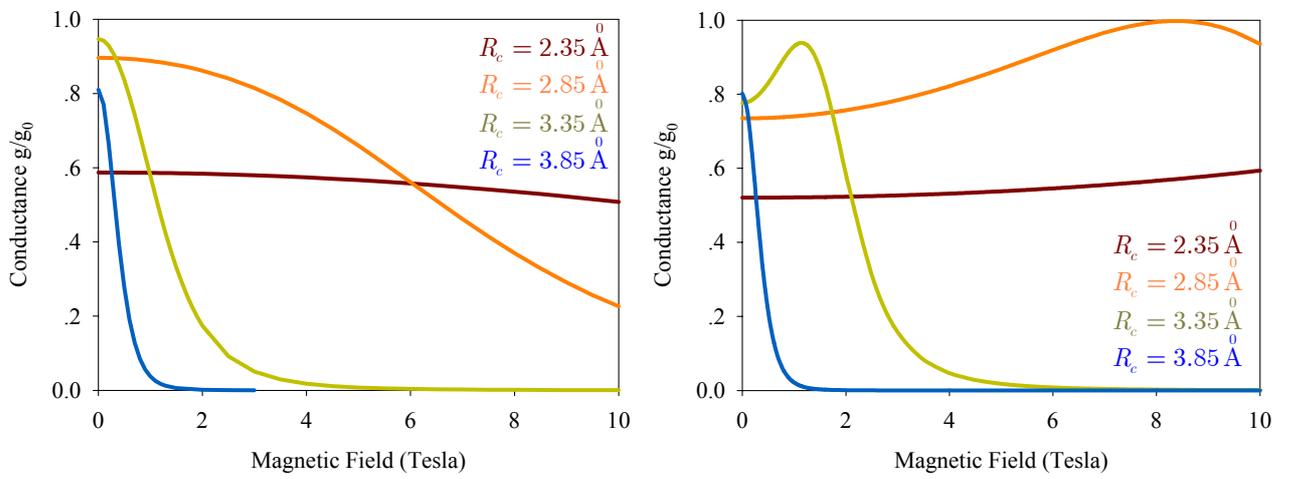